\documentclass[aps,showpacs,prb,twocolumn]{revtex4-1}

\usepackage{graphicx}
\usepackage{dcolumn}
\usepackage{bm}

\begin{document}


\title{Pressure effects on the heavy-fermion antiferromagnet CeAuSb$_2$}%

\author{S. Seo$^1$, V. A. Sidorov$^{2, 3, 4}$, H. Lee$^{2, 5}$, D. Jang$^1$, Z. Fisk$^6$, J. D. Thompson$^2$, T. Park$^1$}
\affiliation{$^1$ Department of Physics, Sungkyunkwan University, Suwon 440-746, Korea \\
$^2$ Los Alamos National Laboratory, Los Alamos, New Mexico 87545, USA\\
$^3$ Institute for High Pressure Physics of Russian Academy of Sciences, RU-142190 Troitsk, Moscow Region, Russia\\
$^4$ Moscow Institute of Physics and Technology, RU-141700 Dolgoprudny, Moscow Region, Russia\\
$^5$ Department of Applied Physics and Geballe Laboratory for Advanced Materials,\\ Stanford University, Stanford, California 94305, USA\\
$^6$ Department of Physics and Astronomy, University of California at Irvine, CA 92697, USA}

\date{\today}

\begin{abstract}
 The {\it f}-electron compound CeAuSb$_2$, which crystallizes in the ZrCuSi$_2$-type tetragonal structure, orders antiferromagnetically between 5 and 6.8~K, where the antiferromagnetic transition temperature \emph{T$_N$} depends on the occupancy of the Au site. Here we report the electrical resistivity and heat capacity of a high-quality crystal CeAuSb$_2$ with $T_N$ of 6.8~K, the highest for this compound. The magnetic transition temperature is initially suppressed with pressure, but is intercepted by a new magnetic state above 2.1~GPa. The new phase shows a dome shape with pressure and coexists with another phase at pressures higher than 4.7~GPa. The electrical resistivity shows a $T^2$ Fermi liquids behavior in the complex magnetic state, and the residual resistivity and the $T^2$ resistivity coefficient increases with pressure, suggesting the possibility of a magnetic quantum critical point at a higher pressure.
\end{abstract}
\maketitle

\section{Introduction}
Cerium-based compounds have attracted attention because they exhibit a variety of interesting phenomena, such as heavy fermion, magnetic, and unconventional superconducting states.~\cite{1,2,3} Especially, magnetically ordered Ce compounds have been model systems for exploring the interplay between magnetism and superconductivity, where the ground state is determined by the balance between Ruderman-Kittel-Kasuya-Yosida (RKKY) and Kondo interactions.~\cite{4} The RKKY interaction with \emph{T$_{RKKY}$}$\sim$£ü\emph{N(E$_F$)J}£ü$^2$ is responsible for a long-ranged magnetic order, while the Kondo interaction with \emph{T$_K$}$\sim$exp(-1/£ü\emph{JN(E$_F$)}£ü) promotes formation of a non-magnetic heavy fermi liquids state. Here \emph{J} is the exchange coupling between the Ce 4f spin and the conduction electrons and \emph{N(E$_F$)} is the density of states at the Fermi level \emph{E$_F$}. The fact that these two competing interactions depend on \emph{JN(E$_F$)}, which can be tuned by non-thermal parameters such as magnetic field, chemical composition or pressure, provides an avenue to control the magnetic transition temperature to $T=0$~K, a quantum-critical point (QCP).~\cite{5,6,7} Non-Fermi liquid (NFL) states that are characterized by a divergence of the effective mass and deviation from a $T^2$ dependent resistivity at low temperatures have been shown to occur due to the abundant quantum fluctuations associated with the QCP.~\cite{5}

CeAuSb$_2$ belongs to the family Ce\emph{T}Sb$_2$ ($T$ = Au, Ag, Ni, Cu, or Pd), a dense Kondo system with pronounced crystalline electric field (CEF) effects.~\cite{10,11,12,13,14} It crystallizes in the ZrCuSi$_2$ type tetragonal structure where the Au layer is contained between two CeSb layers.~\cite{16} Transport and thermodynamic measurements of CeAuSb$_2$ have shown that a long-range antiferromagnetic order develops below 6~K.~\cite{10,17} When a magnetic field is applied along the interlayer direction, Landau-Fermi liquid behavior breaks down at 5.3~Tesla, where the electrical resistivity shows a $T^{1/2}$ dependence and the low-$T$ specific heat $C/T$ diverges logarithmically ($\propto -ln T$), indicating a field-tuned quantum critical point (QCP).~\cite{17} Observation of a small hysteresis at 22~mK under magnetic field, however, suggests a weakly first order nature of the magnetic transition, questioning the origin of the Non-Fermi liquid behavior.~\cite{17} Externally applied pressure, another non-thermal parameter, can control the f-ligand hybridization, thus leading to a pressure-tuned QCP by suppressing the antiferromagnetic phase. Even though earlier pressure work on CeAuSb$_2$ showed that $T_N$ is suppressed to 0~K at 2~GPa,~\cite{10} detailed measurements of physical properties that can reveal quantum critical behavior have yet to be performed under pressure. In addition, $T_N$ of the specimen reported in Ref.~8 was 5.0~K and the residual resistivity ratio (RRR) was small ($\approx 3$), indicating a non-negligible deficiency in the Au site. Because disorder can strongly affect the nature of a quantum critical point,~\cite{vojta10} it is desirable to work on high-quality single crystalline compounds. Here, we report electrical resistivity and specific heat measurements of CeAuSb$_2$, where the occupancy of the Au site is close to 99~\%, $T_N$ is 6.8~K, the highest reported so far, and the RRR is 14, indicating high quality of the measured crystals. Unlike previous work on a crystal with $T_N=5.0$~K, application of pressure does not reduce $T_N$ to 0~K, but rather induces a new, complex magnetic phase at lower temperatures. A $T^2$ Fermi-liquid dependence of the electrical resistivity is observed to the highest measured pressure 5.5~GPa, but the resistivity coefficient $A$ and the residual resistivity $\rho_0$ monotonically increase, suggesting that a magnetic quantum critical point may exist at a higher pressure.

\section{Experiment}
Single crystal CeAuSb$_2$ was synthesized by a self-flux method described in Ref.~16 with Au excess to eliminate deficiency in the Au-site. Electron microprobe analysis using a 5-wavelength dispersive spectrometer was performed on samples to confirm stoichiometric composition. The antiferromagnetic transition temperature strongly depends on the occupancy of the Au site: $T_N$ is 6.8, 6.0, 3.9~K for site occupancies of 99, 93, 89~\%, respectively. For the present study, we have used crystals from a batch that shows an average filling fraction of 99~\% and RRR of 14. Pressure work was performed using a hybrid Be-Cu/NiCrAl clamp-type pressure cell to 3~GPa and a toroidal anvil cell equipped with a boron-epoxy gasket and Teflon capsule up to 5.5~GPa. Pressure in the cells was determined from the pressure dependent superconducting transition temperature of lead using the pressure scale of Eiling and Schilling.~\cite{19} A conventional four-probe technique was used to measure the in-plane electrical resistivity of CeAuSb$_2$ via an LR700 Resistance Bridge in a $^4$He cryostat from 300 K down to 1.2~K. \textit{AC} calorimetry was used to measure the heat capacity,~\cite{yaakov02} where heat is provided by an alternating current of frequency $f$ to a heater attached to the back face of the crystal. A type~E thermocouple was attached to the other face of the crystal to measure the oscillating sample temperature $T_{ac}$ as a $2f$ signal incurred by the oscillating heat input. When the measuring time $\tau = 1/f$ is in an optimum range, i.e., $\tau_1 << \tau << \tau_2$, the oscillating temperature is inversely proportional to heat capacity ($T_{ac} \propto 1/C$). Here the characteristic constants $\tau_1$ and $\tau_2$ are internal sample relaxation and sample-to-bath relaxation times, respectively.

\section{Results and discussion}

Figure~1(a) shows the in-plane electrical resistivity of CeAuSb$_2$ in the vicinity of the antiferromagnetic transition temperature 6.8~K, which is marked by an arrow. As shown in Fig.~1(b), a corresponding anomaly in the specific heat occurs at 6.6~K, a slightly lower temperature than that of the resistivity because of a \textit{dc} offset temperature $T_{dc}$ from the oscillating heat input that is characteristic of the \textit{ac} technique.~\cite{park04} The sharp resistivity drop at $T_N$ and the symmetric shape of the specific heat anomaly are suggestive of a first-order nature of the phase transition. Figure~1(c) shows the higher harmonics of $3f$ and $4f$ of the oscillating sample temperature. The lack of any signature near $T_N$ is consistent with a weakly first-order or a second-order nature of the phase transition because latent heat associated with a first-order phase transition will deform the ideal $2f$ voltage signal and will induce a peak structure at $T_N$ in the higher harmonics.
\begin{figure}[tbp]
\begin{minipage}{18pc}
\includegraphics[width=18pc]{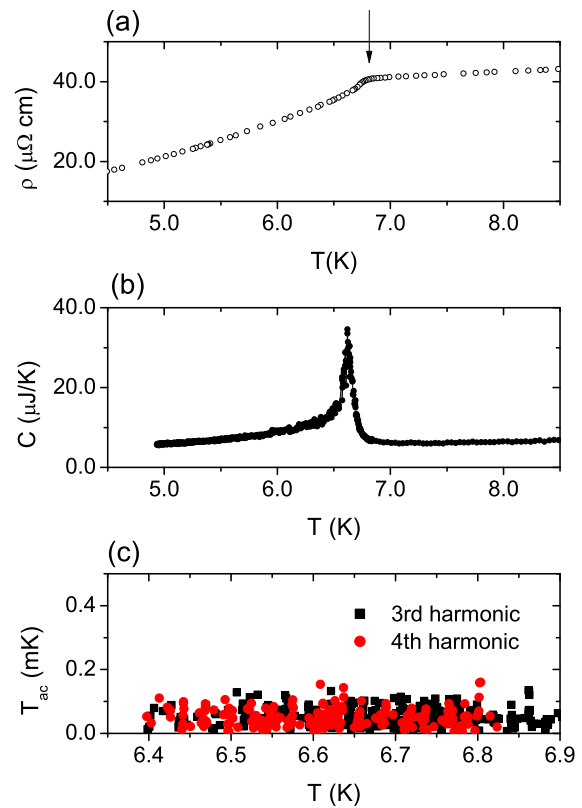}
\end{minipage}\hspace{1pc}%
\caption{\label{label} (a) Temperature dependence of the in-plane electrical resistivity $\rho$ of CeAuSb$_2$ at ambient pressure. The arrow indicates onset of the antiferromagnetic transition at 6.8~K. (b) Specific heat capacity of CeAuSb$_2$. The sharp peak is displaced as a function of temperature relative to the kink in resistivity near $T_N$. (c) Measurements of $3f$ and $4f$ harmonics of $T_{ac}$ while the sample heater is excited at frequency $1f$.}
\end{figure}

Temperature dependence of the in-plane electrical resistivity of CeAuSb$_2$ is plotted at various pressures from 1~bar to 5.5~GPa in Fig.~2(a). The resistivity at room temperature increases with pressure, initially at 3.27 $\mu\Omega \cdot$cm~/~GPa and at a lower rate above 3~GPa. For comparison, we also show the resistivity of LaAuSb$_2$, the non-magnetic analog of CeAuSb$_2$, which linearly decreases with decreasing temperature down to 100~K, but shows an anomaly at lower temperature due to opening of a charge-density-wave gap. The non-magnetic contribution to the resistivity of CeAuSb$_2$ accounts for less than 50~\% of the total resistivity over most of temperature range.
\begin{figure}[tbp]
\begin{minipage}{18pc}
\includegraphics[width=18pc]{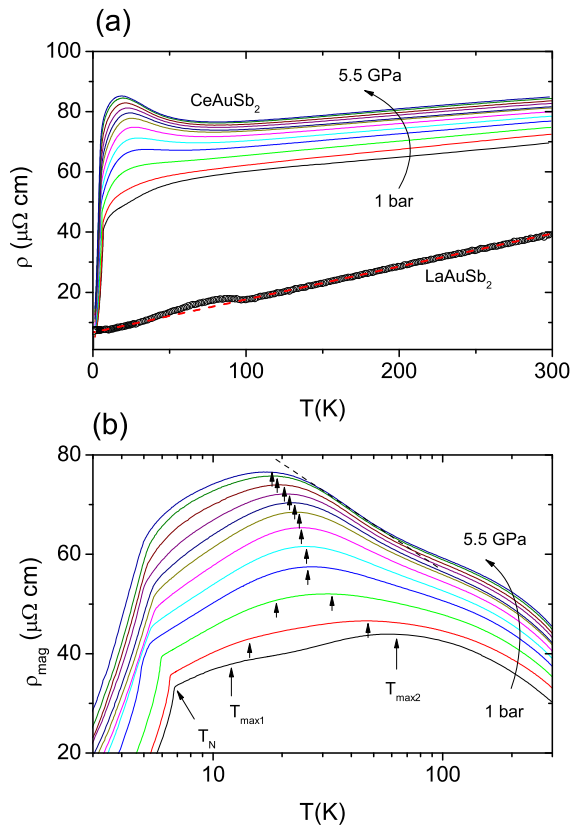}
\end{minipage}\hspace{1pc}%
\caption{\label{label} (a) Temperature dependence of the in-plane electrical resistivity $\rho$ of CeAuSb$_2$ at pressures of 1~bar, 0.8, 1.47, 2.18, 2.72, 3.22, 3.70, 3.86, 4.28, 4.70, 5.14, and 5.50~GPa from the tail to the head of the arrow, respectively. The resistivity of LaAuSb$_2$ is also shown for reference and the dashed line is a linear fit to the resistivity above 100~K. (b) Magnetic contribution to the electrical resistivity $\rho_{mag}$ of CeAuSb$_2$ shown as a function of temperature on a semi-logarithmic scale. Arrows indicate the evolution of peak positions with pressure. The antiferromagnetic transition temperature $T_N$ is also marked by an arrow. The dashed line is a guide to the logarithmic divergence of the $\rho_{mag}$, i.e., $\propto -lnT$, due to Kondo scattering.}
\end{figure}

Figure~2(b) shows the magnetic contribution to resistivity from scattering by Ce $4f$ moments as a function of temperature on a semi-logarithmic scale. The non-magnetic contribution, which is shown as a dashed line in Fig.~2(a), is assumed to be linear down to the lowest measured temperature and independent of pressure because the resistivity change with pressure is expected to be relatively small compared to that of the CeAuSb$_2$. At ambient pressure, the magnetic resistivity $\rho_{mag}$ increases initially with decreasing temperature due to the scattering of itinerant electrons from the Ce $4f$ moments and shows broad peaks near 65~K and 12~K, which are marked as $T_{max2}$ and $T_{max1}$, respectively. With increasing pressure, the two characteristic temperatures evolve oppositely: $T_{max1}$ is enhanced, but $T_{max2}$ is suppressed with pressure. At pressures above 2.18~GPa, the two temperatures merge and the single broad maximum decreases with pressure, where the peak position becomes less clear. The low-temperature peak at $T_{max1}$ may be associated with the emergence of coherence in the lattice of Kondo moments, leading to the formation of a heavy Fermi liquid and decrease in the resistivity. The increase in $T_{max1}$ then reflects enhancement of the hybridization that is typical of Ce-based compounds as a function of pressure. The second energy scale $T_{max2}$ is comparable to that of the crystalline electric field (CEF) splittings. The six-fold multiplet ($J=5/2$) of Ce$^{3+}$ splits into three doublets, with two doublets separated from the ground state by 97 and 145~K, respectively.~\cite{10} The fact that there is only one broad peak near $T_{max2}$(=65~K) at ambient pressure may be due to broadening of these two levels by Kondo scattering on each of the doublets and the relative closeness of the two CEF splittings. With increasing pressure, the interplay between pressure-dependent changes in CEF splitting and hybridization produces a single broad peak as shown in Fig.~2(b) and raises the possibility that wavefunctions of the excited crystal-field levels become admixed into the ground state.

\begin{figure}[tbp]
\begin{minipage}{18pc}
\includegraphics[width=18pc]{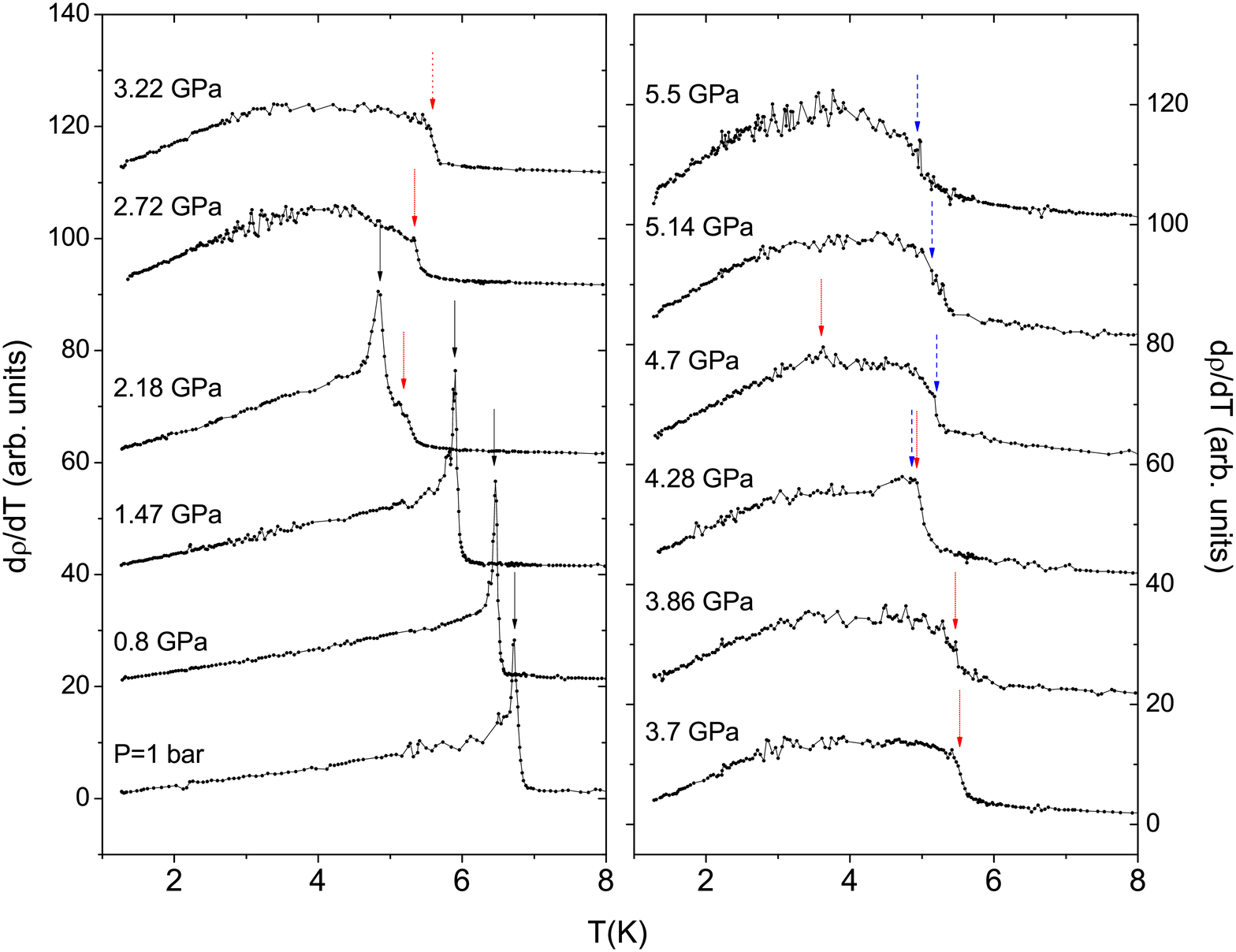}
\end{minipage}\hspace{1pc}%
\caption{\label{label} (a) Temperature derivative of the resistivity at low temperatures. Data at different pressures are shifted rigidly upward for clarity. Arrows mark the characteristic temperatures that are associated with phase transitions: solid, dotted, and dashed arrows for the antiferromagnetic transition, new phase~1, and new phase~2, respectively.}
\end{figure}
Pressure evolution of the antiferromagnetic phase is complex, as indicated in Fig.~2(b). The resistive signature for the antiferromagnetic state is suppressed gradually to lower temperatures with pressure, but survives up to the highest applied pressure of 5.5~GPa. This is more obvious in Fig.~3 where we plot the temperature derivation of the resistivity of CeAuSb$_2$ under pressure. Features in the data enable us to map out a sequence of phase transitions in the low$-T$ magnetic phase diagram. At ambient pressure, the sharpest change in resistivity or the peak in $d\rho/dT$ that corresponds to $T_N$ is gradually suppressed up to 2.18~GPa, where an additional phase transition occurs at a slightly higher temperature (marked as a dotted arrow) and coexists with the original antiferromagnetic phase. With further increasing pressure, the new phase~1 manifested by the appearance of a shoulder in $d\rho/dT$ replaces the original antiferromagnetic phase and its transition temperature shows a maximum at 3.22~GPa. At 4.7~GPa, a shoulder and broad peak indicate the appearance of a new phase~2 that replaces phase~1 at pressures above 5.14~GPa. A similarily complex sequence of magnetic phases at low temperatures and high pressures has been reported in other heavy fermion antiferromagnetic compounds such as U$_2$Zn$_{17}$, CeZnSb$_2$, and CeNiSb$_3$.~\cite{30, park05, sidorov05}

\begin{figure}[tbp]
\includegraphics[width=18pc]{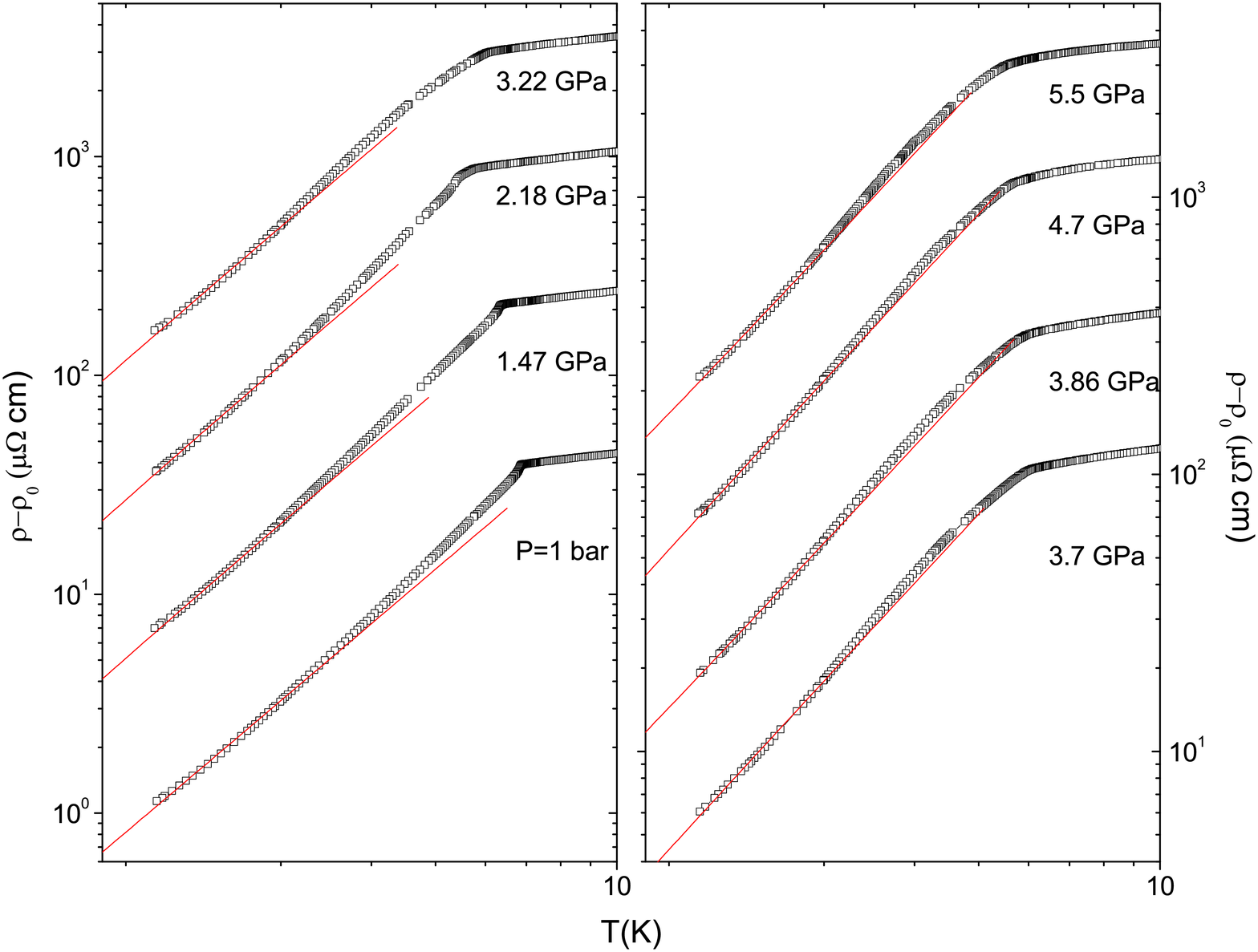}\hspace{2pc}%
\begin{minipage}[b]{18pc}\caption{\label{label}Temperature dependence of the in-plane electrical resistivity of CeAuSb$_2$ after subtracting the residual resistivity $\rho_0$ is plotted on logarithmic scales. Solid lines are least-squares fits of $\rho-\rho_0=AT^2$. The resistivity curves obtained at applied pressures are shifted rigidly upward from 1~bar data for clarity.}
\end{minipage}
\end{figure}
Figure~4 shows the low-temperature resistivity as a function of temperature in a log-log plot for several pressures. The solid lines are least-squares fit of $\rho-\rho_0=AT^n$. Here $\rho_0$ is a residual resistivity and the temperature exponent $n$ is the slope of the curves. Independent of pressure, $n=2$ best explains the low-temperature resistivity, indicating the absence of significant magnon scattering and the dominance of Landau-Fermi liquid behavior in the various magnetic phases. At ambient pressure, the Sommerfeld coefficient $\gamma$ obtained from the $T^2$ coefficient $A$ (=0.56~$\mu \Omega \cdot$cm$\cdot$K$^{-2}$) is 236~mJ/mol$\cdot$K$^{2}$, where we have used the Kadowaki-Woods ratio R$_{KW}$=A/$\gamma$$^2$=1.0$\times$10$^{-5}$$\mu$$\Omega$$\cdot$cm$\cdot$mol$^2$$\cdot$\emph{K}$^2$$\cdot$mJ$^{-2}$ (Ref.~23). This large value is comparable to the previous specific heat measurement that showed 50~\% of $Rln2$ entropy recovered at the magnetic ordering temperature,~\cite{10} consistent with the existence of a heavy Fermi liquid in the antiferromagnetic state.
\begin{figure}[tbp]
\begin{minipage}{18pc}
\includegraphics[width=18pc]{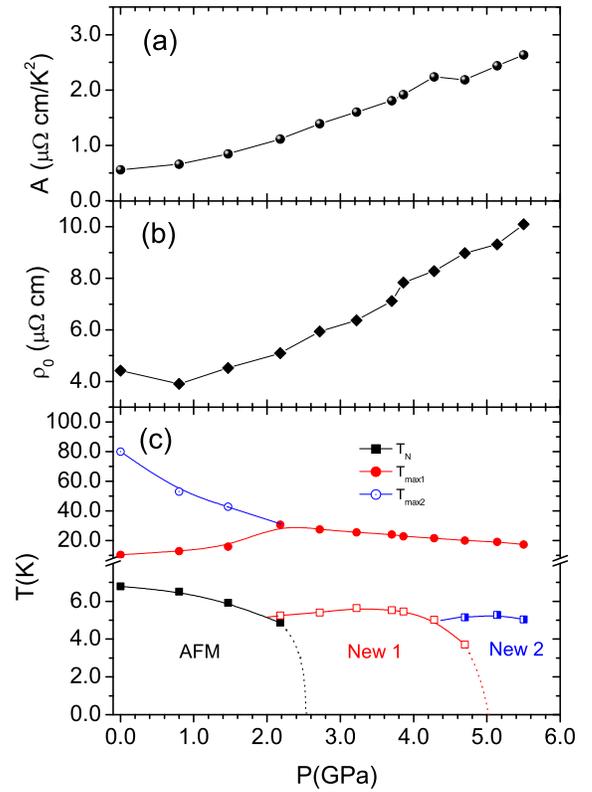}
\end{minipage}\hspace{1pc}%
\caption{\label{label} Pressure dependence of the $T^2$ coefficient $A$ and the residual resistance $\rho$$_0$ are shown in (a) and (b), respectively. (c) Temperature-pressure phase diagram of CeAuSb$_2$ constructed from resistivity measurements. $T_{max1}$ and $T_{max2}$ are maximal temperatures in $\rho$ that are associated with Kondo coherence and CEF effects, respectively. At low temperatures, the N$\acute{e}$el temperature and new phases 1 and~2 are indicated by solid, open, and half-filled square symbols, respectively. Dashed curves are suggested extrapolations of phase boundaries.}
\end{figure}

The pressure dependence of the $T^2$ coefficient $A$ and residual resistivity $\rho_0$ of CeAuSb$_2$ is plotted in Fig.~5(a) and 5(b). The coefficient $A$ gradually increases with pressure and reaches 2.64~$\mu \Omega \cdot$cm$\cdot$K$^{-2}$ at 5.5~GP. Assuming the Kadowaki-Woods ratio is pressure independent, the factor of 4.7 increase in $A$ implies that the effective electron mass is enhanced by a factor of 2.2. Likewise, the residual resistivity increases from 4.42 at 1~bar to 10.09~$\mu\Omega\cdot$cm at 5.5~GPa, a relative increase similar to that of effective mass.

A monotonic increase in the $A$ coefficient and $\rho_0$ is found in several heavy-fermion antiferromagnets as they are tuned toward a quantum critical point.~
\cite{5} As shown in Fig.~5(c), however, the pressure-dependent evolution of magnetic phases in this high-quality single crystal is not monotonic, but is very complex. The N$\acute{e}$el temperature $T_N$ of the initial AFM phase (solid squares) decreases gradually with pressure, but its approach to $T=0$~K is interrupted by a new phase~1 (open squares) at 2.18~GPa, whose transition to $T=0$~K also appears to fail due to the formation of new phase~2. These observations lead to the possibility of a series of 'failed' quantum phase transitions that are responsible for the overall increase in $A$ and $\rho_0$. The new phases~1 and~2 are energetically favored relative to the potential quantum disorder state. The disappearance of the AFM phase and appearance of the new phase~1 seems coincident with the merger of the two characteristic scales $T_{max1}$ and $T_{max2}$. As mentioned above, it seems possible that the ground state wavefunction becomes admixed with those of higher lying crystal fields near this pressure. In this regard, it is interesting that even at higher pressures the single broad maximum moves to lower temperatures, as shown in Fig.~5(c), which is opposite to the pressure dependence found commonly in Ce-based Kondo lattice compounds where the competition between RKKY and Kondo interactions dominates the pressure response.~\cite{jaccard99} If this unusual pressure dependence on CeAuSb$_2$ is being dominated by a decrease in CEF splitting, with decreasing volume,~\cite{oomi96}, then the additional admixture of CEF and groundstate wavefunctions could contribute to the energetic stability of pressure-induced phases~1 and~2. At pressures above those reached in these experiments, magnetic order must disappear as the f-ligand hybridization becomes sufficiently strong. Whether superconductivity will appear at the magnetic-non-magnetic boundary, as it does in several Ce-based heavy fermion systems,~\cite{monthoux07} remains to be seen.

\section{Conclusion}

In summary, we have measured the electrical resistivity of a high-quality, nearly stoichiometric crystal CeAuSb$_2$ up to 5.5~GPa. The magnetic phase diagram of CeAuSb$_2$ under pressure is complex, probably reflecting the interplay of the pressure dependence of RKKY, Kondo and CEF effects. $T_N$ is suppressed with pressure, coexists with a new phase at 2.1~GPa, and is replaced by new phase~1 at 2.7~GPa. The new phase~1 has a dome shape and is followed by a new phase~2 above 4.7~GPa. The similarity of transition temperatures and the evolution of $d\rho/dT$ suggests that the two new phases are magnetic in character. A $T^2$ dependence in the low-temperature resistivity reflects the dominance of Landau-Fermi liquid behavior across the $T-P$ phase diagram. The gradual enhancement of the $T^2$ coefficient $A$ and residual resistivity $\rho_0$ up to the highest measured pressure 5.5~GPa suggests that a magnetic quantum critical point may exist at higher pressure.

This work was supported by NRF grant funded by Korea government (MEST) (No.~2010-002672, 2011-0021645, \& 220-2011-1-C00014). Work at Los Alamos was performed under the auspices of the U. S. Department of Energy/Office of Science and supported in part by the Los Alamos LDRD program. ZF acknowledges support from NSF-DMR-0801253. VAS is supported in part from Russian Foundation for Basic Research (RFBR grant 12-02-00376).


\end{document}